\documentclass[lettersize,journal]{IEEEtran}
\usepackage{amsmath,amsfonts}
\usepackage{algorithm}
\usepackage{array}
\usepackage[caption=false,font=normalsize,labelfont=sf,textfont=sf]{subfig}
\usepackage{textcomp}
\usepackage{stfloats}
\usepackage{url}
\usepackage{verbatim}
\usepackage{graphicx}
\usepackage{cite}
\usepackage{algorithm}
\usepackage{algpseudocode}
\usepackage{algorithmicx}

\begin{document}
%

\title{Real Time Vehicle Identification: \\A Synchronous-Transmission Based Approach}








\author{Chandra Shekhar, Manish Kausik H, Sudipta Saha
\thanks{The authors are with the Computer Science and Engineering, School of Electrical Sciences, Indian Institute of Technology Bhubaneswar, India. Email: \{cs13, mkh10, sudipta\}@iitbbs.ac.in}
\thanks{Manuscript received August XX, 2023.}}

\markboth{Journal of \LaTeX\ Class Files,~Vol.~14, No.~8, August~2023}%
{Shell \MakeLowercase{\textit{et al.}}: A Sample Article Using IEEEtran.cls for IEEE Journals}



%


\maketitle



%
\IEEEpeerreviewmaketitle

\begin{abstract}
Identification of the vehicles passing over the roads is a very important component of traffic monitoring/surveillance. There have been many attempts to design and develop efficient strategies to carry out the job. However, from the point of view of practical usefulness and real-time operation, most of them do not score well. In the current work, we perceive the problem as an efficient real-time communication and data-sharing between the units in charge of recording the identities of the vehicles, i.e., \textit{Vehicle Recorders} (VR), and the \textit{Vehicles} (VE). We propose a strategy to address the issue with the help of \textit{Synchronous-Transmission} (ST), which is a newer paradigm of communication compared to the traditional paradigm based on \textit{Asynchronous-Transmission} (AT). First, we theoretically show that the presence of the physical layer phenomena called \textit{Capture-Effect} in ST brings a significant benefit. Next, we also implement the strategy in a well-known IoT-Operating System \textit{Contiki} and compare its performance with the existing best-known strategy.

\end{abstract}

\vspace{-0.3cm}
\section{Introduction}
\label{sec:intro}

Identification of the vehicles moving over roads is one of the very significant components of an \textit{Intelligent-Transportation System} \cite{ITS}. It plays a very important role in crime investigation, vehicle tracking \cite{vehicle-tracking}, executing vehicle-specific action on demand \cite{Rajeshwari}, etc. There have been various works to accomplish the goal. Outcomes from these works, although, can successfully detect and identify the vehicles dynamically upto a certain extent, they bear a number of drawbacks and limitations as stated below.

A large class of works uses \textit{Computer-Vision} \cite{realtime-arya} along with \textit{OCR} techniques \cite{realtime-ocr} to read the \textit{Vehicle Registration Numbers} (VRN) directly from the number plates of the vehicles. However, it requires \textbf{high visibility} which may not be feasible in many situations, e.g., foggy or rainy conditions, night-time or low-light situations, etc. Besides, these strategies naturally need both \textbf{constant supply of power}, as well as uninterrupted \textbf{cloud connectivity} to support heavy computation. These constraints make them highly inappropriate for widespread use in many demanding places across traffic systems.

Another class of solutions employs passive/active RFID tags \cite{zhoulicense, RFID-new} for rapid identification of the vehicles. RFID technology can cope with the issue of visibility, as well as, can be quite cost-effective too. However, unfortunately, these solutions heavily lack \textbf{Scalability}. In particular, RFID-based solutions can detect and identify vehicles quite accurately and efficiently when they appear in a highly systematic manner, e.g., vehicles passing through the toll plaza or entering a car-parking region one-by-one in a quite restricted setting. However, in a generalized transportation scenario, there can be a wide variation in the number and the speed of the vehicles passing together. Unfortunately, RFID-based solutions, due to a lack of efficient and scalable recording mechanisms, fail to cope up with such scenarios. Computer-vision-based strategies also have similar limitations on the same ground.



Thus, despite being highly demanding, real-time identification of the vehicles is still a quite challenging task. A solution needs to balance a certain set of conflicting goals as follows. (a) To get installed over multiple different locations over a wide area the solution needs to be both {cost-effective} and \textit{simple}. (b) The solution should be able to manage an unrestricted number of vehicles where only a limited number of them will be present at a time. (c) Finally, the solution has to ensure high accuracy in dynamically identifying the vehicles.

\begin{figure}[htbp]
\begin{center}
\includegraphics[angle=0,width=0.5\textwidth]{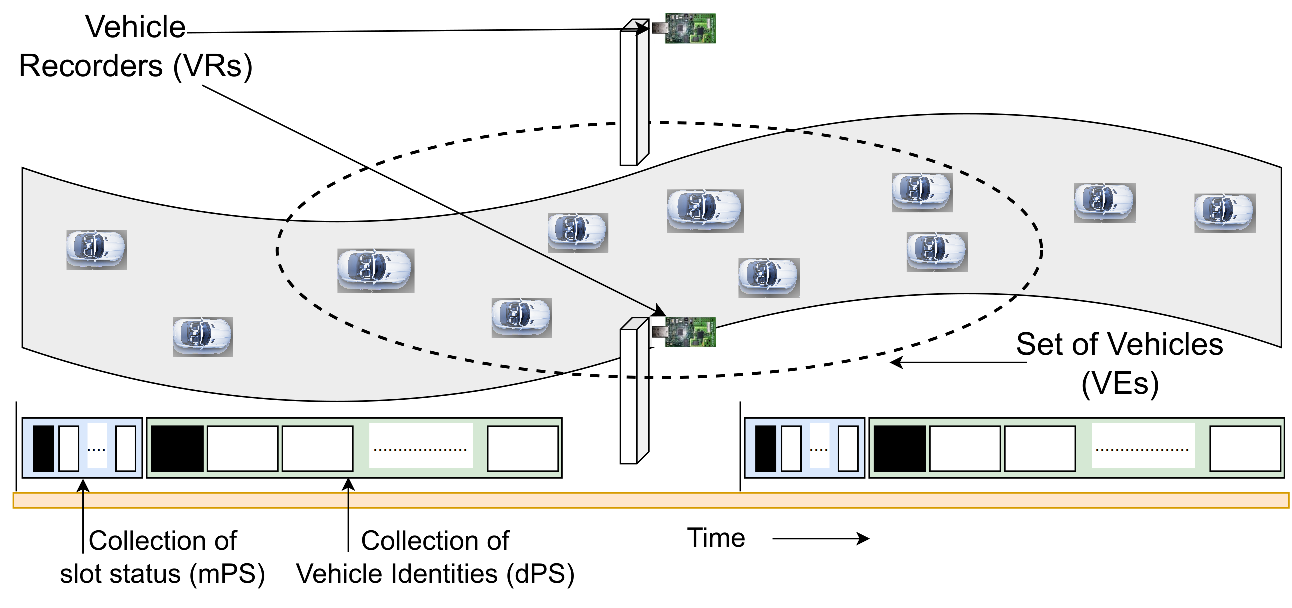}
\end{center}
\caption{ A possible scenario with two VRs and multiple VEs and time diagram showing the mPS and dPS.} 
\label{fig:identification}
\end{figure}

In the current work, we view the problem of vehicle identification as an issue of carrying out a fruitful communication between two sets of entities (a) \textit{A semi-defined set of} \textit{Vehicles} (VE) \textit{whose cardinality is very large but at a time only a limited no of them are present}. (b) \textit{A set of} \textit{Vehicle Recorders} (VR) in charge of recording the VRNs of the vehicles. \textit{In particular, we perceive it as a problem of fast/instantaneous data-sharing to convey the VRNs of the VEs to the VRs as soon as they come into contact with each other at the discrete recording locations.} Fig \ref {fig:identification} shows the setting composed of VEs and the VRs.

Existing works in a similar direction mostly use traditional \textit{Asynchronous-Transmission} (AT) based communication mechanism where the MAC layer is constructed using CSMA/CA which resolves most of the issues related to collisions among the packets with the help of random delays. Highly dynamic situations with diversity in the density and the speed of the vehicles make an AT-based strategy very inappropriate for recording the identities of the VEs by the VRs in real-time. In contrast, in the current work, we exploit recent advances in \textit{Synchronous-Transmission}(ST)-based communication to address this issue. In particular, we define an IoT-based lightweight and low-cost framework for carrying out the job in real-time.

A stringent limitation of most of the existing works in this direction is their assumption that the set of possible vehicles that may pass through the recording location of the road is quite well-defined with a restricted cardinality. A number of the works hence, employ TDMA slots \cite{plos-one} and assume that a vehicle would respond at a specific TDMA slot as per their identification number (referred to as \textit{Vehicle Registration Number} Or VRN) and get recorded accordingly. However, such assumption is far away from reality. Although at a time a limited number of vehicles would be passing by together, the total set of possible vehicles to be recorded can be extremely large. A natural solution in such a situation is to use TDMA with \textit{hashing}. However, this results in the wastage of a number of slots due to collision among multiple packets transmitted from multiple VEs. In our approach, we show that the use of ST, by virtue of the physical layer phenomena called \textit{Capture-Effect}\cite{capture}, can substantially reduce such wastage and hence, produce significantly better results compared to the existing strategies.

The contributions of the paper are as follows.

\begin{itemize}
    \item We propose a lightweight and low-cost communication-based framework for real-time identification of vehicles passing over the roads.
    \item We demonstrate how ST can be exploited to enhance the capability of TDMA along with hashing for dynamically recording the identities of the vehicles.
    \item We provide a theoretical analysis of the proposed mechanism, as well as implement it in \textit{Contiki OS} for TelosB devices. Furthermore, we also compare its performance with the state-of-the-art solution for vehicle identification. 
\end{itemize}

The rest of the paper is organized as follows. Sec. \ref{sec:design} demonstrates the design of the proposed strategy in detail. Sec. \ref{sec:theory} provides a theoretical study of the same and finally, Sec. \ref{sec:evaluation} provides a simulation-based evaluation study.




\section{Design}
\label{sec:design}

Each of the VRs and VEs is supposed to be equipped with a low-cost low-power IoT-device capable of communicating with each other using some suitable technology, e.g., LoRa, 802.15.4, or BLE. In this work, we particularly use 802.15.4. We consider the problem in hand as an issue of efficient and real-time \textit{many-to-many} communication between the number of VEs and the set of one or more VRs under the setting shown in Fig. \ref{fig:identification}. In other words, its assumed that one or more VRs are installed at a certain location by the side of a road through which the VEs would pass. The VEs are assumed to be under one-hop communication range with the VRs for a short amount of time.


We assume a reactive communication setting where a VR queries (probes) the set of VEs and records their VRNs from the corresponding replies. However, it is quite vivid that since there may be a quite large number of VEs passing by together, solving the massive instantaneous many-to-one data-sharing, i.e., VRNs from many VEs to one VR, for each of the VRs, is a real challenge.

Traditional CSMA/CA-based solutions are not suitable for addressing this issue as they incur collision among packets and introduce random waiting to resolve collision which substantially delays the overall process. In this work, we take the help of ST-based communication where the nodes carry out the tasks in a highly coordinated manner through tight time-synchronization among each other. We adopt an existing ST-based protocol \textit{PacketSync} \cite{packetsync} for this purpose. PacketSync is first briefly described below for completeness.


\begin{figure}[htbp]
\begin{center}
\includegraphics[angle=0,width=0.4\textwidth]{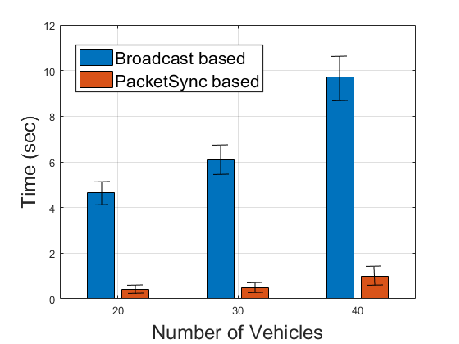}
\end{center}
\caption{Comparison between AT and ST based strategy for simple TDMA based single-hop many-to-one data-collection}
\label{fig:comparison}
\end{figure}

\textbf{PacketSync}: PacketSync \cite{packetsync} is an ST-based many-to-one data-sharing protocol that can efficiently collect data from a number of source nodes to a designated sink node connected in a single-hop (directly). The sink node first sends a short probe message which triggers the source nodes to emit their packets as per a given order. The order is either decided globally or is mentioned on-the-fly in the probe packets. Note that the same task can be also executed using AT-based sequential broadcast by the source nodes as per a pre-decided TDMA schedule. Fig. \ref{fig:comparison} shows a comparison of the performance between AT and ST based realizations. ST-based one, i.e., PacketSync, due its compact form factor, performs several times better than the other in terms of the time requirements (Latency). However, PacketSync assumes a quite defined and limited number of source nodes which is not a realistic scenario in the current problem. In the following, we show the design of our strategy VSync which resolves these issues by employing PacketSync as a base.

\subsection{Vehicle Recording}

The work \cite{plos-one} observes that, when TDMA is combined with hashing, a considerable inefficiency arises due to wastage of time in the \textit{unused-slots}, i.e., those slots which are opted by none of the VEs. It proposes to mark these unused slots by asking the VEs to first quickly send a set of dummy packets through the same hashing strategy. These dummy packets are several times shorter than the data packets bearing the VRNs of the VEs. To save time, the unused slots are then removed and the rest of the slots are re-numbered based on a mapping function which is then conveyed to the VEs through another probe. The VEs respond in the next round with their VRNs as per the prescribed mapping. The strategy is referred to as \textit{Reservation to Cancel Ideal dynamic frame slots} (RTCI) \cite{plos-one}.

 
However, the work \cite{plos-one} demonstrates only the proof of the concept through Matlab-based simulation. It does not show how it can be truly realized in devices. Minute observation reveals that because of the mobility of the VEs, the characteristics of the hashing function sensed through the transmission of the dummy packets can get substantially changed by the time the VEs actually communicate their VRNs. Thus, the proposed solution is useful only when it executes very fast so that the scenario does not get altered. This is quite hard to achieve using traditional AT-based techniques because of the use of CSMA/CA in the MAC layer. 

\subsection{VSync}

In this work, we propose an ST-based strategy VSync to resolve the above-mentioned issues. It builds an efficient framework for collecting the VRNs using two consecutive rounds of \textit{PacketSync}. Extraction of the characteristics of the hashing function using dummy packets is achieved in VSync through ST in a very compact form. In addition, ST-based approach in VSycn enables it to exploit the virtue of CE which. In RTCI, the time savings come only through the avoidance of unused slots. However, the use of CE also helps VSync to exploit the collision slots. In particular, when multiple VEs opt for a single slot, due to CE with a high chance the strongest signal wins. 

Algorithm \ref{algo:VR} and Algorithm \ref{algo:Vehicle} provide the sketch of the proposed strategy in a self-descriptive fashion for the VRs and the VEs, respectively. Each iteration of VSync is preceded by a short Glossy round (for around 15 ms) for time-synchronization of the VEs and VRs. It is followed by two consecutive rounds of PacketSync. The first round (referred to as mPS) is an extremely short one and is used for discovering unused slots. The second round (referred to as dPS) is used for collecting the VRNs in a compact manner. The slots in dPS are renumbered as per the information collected during mPS. Fig. \ref{fig:identification} (bottom) depicts a time-diagram of mPS and dPS.

\textbf{Use of Conflicting-slots:} In ST, time-synchronized communication along with appropriate differences in the signal strength enables the receivers to perfectly receive the packet having the highest signal strength even when multiple packets appear together. CE in VSync, thus, enables the VRs to receive at least one VRN in most of the conflicting slots, i.e., when multiple VEs transmit their VRNs at the same slot. This substantially improves the outcome from VSync compared to the strategy proposed in \cite{plos-one}. In the following, we theoretically demonstrate this special issue in detail.

                

\begin{algorithm}
\small
\scriptsize
\caption{{ \small \textsc{Algorithm for Vehicle-Recorder (VR) }}}\label{algo:VR}
\begin{algorithmic}[1]
\State $DB$ = [] \Comment{Database of Vehicles}
\State Initialize list $m$

\For{every iteration of VSync}
    
\textbf{\underline{{(mPS)}}}\vspace{1mm}
    \State Send a probe to VEs
    \For{every iteration $S_{j}$ in reception phase}
        \If{Packet has valid $CRC$}
            \State $j=j+1;$
            \State Append $S_{j}$ in the list $m$
        \EndIf
    \EndFor
    

\textbf{\underline{{(dPS)}}}\vspace{1mm}

    \State Send Probe to the VEs, Include $m$ in the probe
    \For{Every Slot $P \gets 1$ to $j$}

    \State $P \gets$ Receive Packet from Radio
             \If{If $P_{j}$ has valid $CRC$}
                 \State Append ($VRN$) in DB
                 \State Send Acknowledgement
             \EndIf
    \EndFor
\EndFor
\end{algorithmic}
\end{algorithm}


\begin{algorithm}
\small
\scriptsize
\caption{{\small \textsc{Algorithm for Vehicle (VE)}}}\label{algo:Vehicle}
\begin{algorithmic}[1]
\State Convert the $VRN$ to an integer number $Y$ 
\For{every iteration of VSync}

\textbf{\underline{{(mPS)}}}\vspace{1mm}

    \State Receive the probe of mPS
    \State Calculate the slot number $j=hash(Y)$
    \State Send a dummy small packet at the jth slot
    \State Wait mPS to get complete
    
\textbf{\underline{{(dPS)}}}\vspace{1mm}

    \State Receive the probe of dPS
    \State Read $m$ and calculate its own slot number $t$
    \State Communicate $VRN$ at the slot $t$
    \State Wait dPS to get complete
\EndFor
\end{algorithmic}
\end{algorithm}

\section{Theory}
\label{sec:theory}

Let there be $L$ TDMA slots and $N$ VEs. Under random selection of the slots by an efficient hash function, the probability that a slot will be selected is $p=1/L$. The number of vehicles that place their packet in a given slot follows the binomial distribution with parameters $N$ and $p$. Success for a vehicle to uniquely select a slot depends on the selection by others. The probability that there is exactly one packet placed in a given slot is $P_{1} = \binom{N}{1} \frac{1}{L} (1-\frac{1}{L})^{N-1}$. Thus, the probability that a slot is idle (i.e., none of the VEs select it) is: $P_{0} = \binom{N}{0} (1-\frac{1}{L})^N$. Furthermore, chance of collision is thus, $P_{coll} = 1 - P_{1} - P_{0}$.

A VR can successfully receive a packet in a TDMA slot when (a) Only one packet is placed by only one of the VEs in a slot, or (b) more than one VE attempts to transmit a packet in the slot, but due to CE, only one of the packets could successfully be placed. The dynamics of CE have been studied in many prior works \cite{capture}. We use the parameter $\rho$ to model the probability of successful CE. The throughput ($S$) of the system can be calculated as: $S = \rho P_{coll} + P_{1}$.
Let us consider $L$ to be the number of slots used for TDMA. $S$ can be maximized by setting $\frac{dS}{dL} = 0$ which gives us $L_{optimal} = \rho + (1-\rho)N$. Considering a large number of vehicles, the best throughput that can be achieved for a given estimate of CE ($\rho$) is as follows,
\begin{equation}
    \label{eqn:optimalS}
    S_{optimal} = \rho + (1-\rho)e^{\frac{-1}{1-\rho}}.
\end{equation}

The basic TDMA and hashing-based approach is referred to as \textit{Dynamic Frame Slotted Aloha} (DFSA) \cite{plos-one}. The throughput of DFSA (without any role of CE) can be computed as follows.
\begin{equation}
    \label{eqn:dfsa_S}
    S_{DFSA} = \frac{N}{L}(1-\frac{1}{L})^{N-1}
\end{equation}

Although $N$ is considered to be large, asymptotic convergence of the equation happens even when $N$ crosses 10. The existence of CE, thus, found to be substantially improving the throughput of DFSA as shown in (\ref{eqn:optimalS}) (relationship plotted in \label{fig:srhoplot}) which justifies the design principle of VSync. In contrast, the throughput obtained from the strategy RTCI in \cite{plos-one} can be modeled as $S_{RTCI}=\frac{L_{succ}}{L - L P_0 +L_{ovr}}$ where, $L_{succ}=L (P_1)$. 
\begin{equation}
    \label{eqn:RTCI_S}
    S_{RTCI} = \frac{N(1-p)^{N-1}}{1+L(1-(1-p)^{N}}
\end{equation}
Considering the removal of the idle part in VSync, the throughput can be computed as similar to RTCI where $L_{succ}=L (P_1 + \rho P_{coll})$. $L_{ovr}$ denotes the associated overhead of mPS. Note that, by virtue of compact ST-based implementation, $L_{ovr}$ is fundamentally negligible, and hence we do not consider it further in the calculation. Throughput in VSync can be derived as follows (ignoring the overhead).

\begin{equation}
    \label{eqn:VSync_S}
    S_{VSync} = \frac{(\rho+(1-\rho)Np(1-p)^{N-1} - \rho (1-p)^{N})L}{L(1-(1-p)^N)+1}.
\end{equation}

\begin{figure}[htbp]
\begin{center}
\includegraphics[angle=0,width=0.42\textwidth]{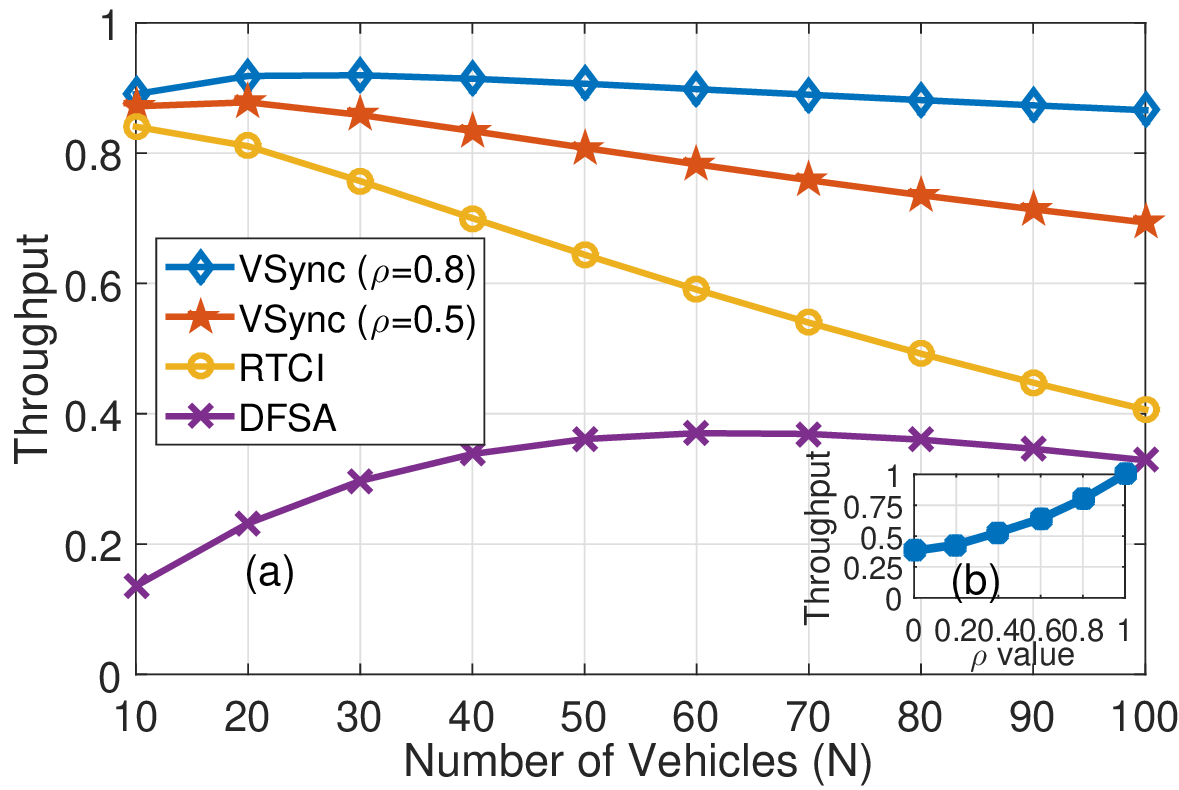}
\end{center}
\caption{Degradation of throughput with increase in the number of VEs in DFSA, RTCI, and VSync for two different values of $\rho$.} \label{fig:theory}
\end{figure}

Fig. \ref{fig:theory}(a) graphically compares the throughput obtained from DFSA (\ref{eqn:dfsa_S}), RTCI (\ref{eqn:RTCI_S}) and VSync (\ref{eqn:VSync_S}). For VSync we consider two values of $\rho$, 0.5 and 0.8 \cite{capture}. The effect of $\rho$ on throughput is explicitly shown in Fig. \ref{fig:theory}(b) (based on (\ref{eqn:optimalS})). It can be seen that with the increase in the number of vehicles throughput of VSync remains considerably higher compared to the other approaches.

\begin{figure}[htbp]
\begin{center}
\includegraphics[angle=0,width=0.4\textwidth]{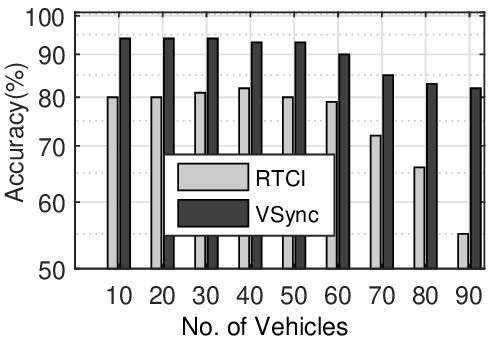}
\end{center}
\caption{Shows the comparison of RTCI and VSync in terms of Accuracy} 
\vspace{-0.5cm}\label{fig:comparision}
\end{figure}
\section{Evaluation}
\label{sec:evaluation}

We implement VSync in Contiki Operating System for TelosB devices. The strategy is simulated in the default simulator of Contiki, i.e., \textit{Cooja} with a single VR and multiple VEs passing near the VR as shown in Fig. \ref{fig:identification}. Cooja does not support mobility. We made the necessary with the Java-based GUI of Cooja to incorporate the mobility of the VEs to perfectly simulate the scenario. We compare the performance of VSync with RTCI in terms of the metric \textit{Accuracy} which is defined as the ratio of the \textit{no of VEs whose VRNs the VR could correctly detect} and \textit{the total no of VEs present within the communication range of the VR}. It is presented (in \% form) as an average over at least 5000 iterations of VSync. The performance is compared with that of RTCI as presented in \cite{plos-one}. It can be seen from Fig. \ref{fig:comparision} that appropriate exploitation of CE in VSync enables it to perform significantly better than RTCI, especially with the increase in the no of VEs.




\section{Conclusion}
\label{sec:conclusion}

We propose an ST based strategy for real-time identification of the vehicles passing over the roads. We theoretically show the presence of the physical layer phenomena called \textit{Capture-Effect} under ST significantly benefits the proposed strategy. Finally, we implement the proposed strategy in Contiki OS, and through a simulation-based study, we show that it performs significantly better compared to the state-of-the-art strategy.


\bibliographystyle{abbrv}
\bibliography{sample.bib}

\end{document}